# Fast two-snapshot structured illumination for temporal focusing microscopy with enhanced axial resolution

Yunlong MENG, Wei LIN, Chenglin LI, Shih-Chi CHEN*

Department of Mechanical and Automation Engineering, The Chinese University of Hong Kong, N.T., Hong Kong

*sschen@mae.cuhk.edu.hk

## ABSTRACT

We present a new two-snapshot structured light illumination (SLI) reconstruction algorithm for fast image acquisition. The new algorithm, which only requires two mutually $\pi$ phase-shifted raw structured images, is implemented on a custom-built temporal focusing fluorescence microscope (TFFM) to enhance its axial resolution via a digital micromirror device (DMD). First, the orientation of the modulated sinusoidal fringe patterns is automatically identified via spatial frequency vector detection. Subsequently, the modulated in-focal-plane images are obtained via rotation and subtraction. Lastly, a parallel amplitude demodulation method, derived based on Hilbert transform, is applied to complete the decoding processes. To demonstrate the new SLI algorithm, a TFFM is custom-constructed, where a DMD replaces the generic blazed grating in the system and simultaneously functions as a diffraction grating and a programmable binary mask, generating arbitrary fringe patterns. The experimental results show promising depth-discrimination capability with an axial resolution enhancement factor of 1.25, which matches well with the theoretical estimation, i.e, 1.27. Imaging experiments on pollen grain samples have been performed. The results indicate that the two-snapshot algorithm presents comparable contrast reconstruction and optical cross-sectioning capability with those adopting the conventional root-mean-square (RMS) reconstruction method. The two-snapshot method can be readily applied to any sinusoidally modulated illumination systems to realize high-speed 3D imaging as less frames are required for each in-focal-plane image restoration, i.e., the image acquisition speed is improved by 2.5 times for any two-photon systems.

## 1. INTRODUCTION

Two-photon excitation (TPE) microscopy has been widely used in the field of bio-imaging and noninvasive diagnosis since its first introduction in the early 1990s [1]. Importantly, TPE microscopy presents distinctive advantages in deep tissue and *in vivo* imaging due to its long (near-infrared) excitation wavelength and low photobleaching/phototoxicity effect respectively [2]. A generic TPE system employs a scanning system that raster-scans the laser focus through a specimen; the emissions from each pixel are then collected to form an optical cross-section. To increase the imaging speed, i.e., video rate or higher, one direct approach is to use high speed scanners, e.g., resonant galvanometric scanners, polygonal scanners, and acousto-optic deflectors (AODs) [3–5]. However, in these systems high scanning speeds are achieved at the expense of pixel dwell time and may compromise the signal-to-noise ratio (SNR). To address the issue, one may excite or illuminate multiple pixels in the specimen simultaneously. Good examples include temporal focusing fluorescence microscopy (TFFM), and line-scan or multifocal multi-photon microscope [6–14]. Among the aforementioned methods, TFFM has a higher degree of parallelization, i.e., higher speed. TFFM is realized by spatially and temporally focusing a femtosecond laser in the focal region [15]. Accordingly, the excitation speed can be as fast as the laser repetition rate [16]; and the imaging speed is practically limited by the speed of cameras, i.e., 1 – 10 kHz.

Compared with a point-scanning system, the axial resolution of the wide-field TFFM is slightly compromised [17,18], which prevents it from broad adoption and commercialization. Structured light illumination (SLI) has been applied in TFFM to improve its optical cross-sectioning capability [19]. In addition, SLI algorithm also helps to suppress the effect of tissue scattering by acting as a virtual pinhole that computationally filters out the scattered photons outside the focal plane based on a set of raw structured images [20]. However, for any two-photon imaging system, at least five equal phase-stepped images are required to remove the sinusoidal modulation terms and obtain the in-focus image when the conventional root-mean-square (RMS) method is applied [21, 22], which compromises the speed of TFFM. One way to reduce the required raw structured images is to apply the HiLo technique [23–25], which applies a high-pass filer to a uniform image, and a low-pass filter to a structured image to form a fused HiLo image that contains all frequencies that can be collected by the image system. However, the optimal filter cut-off frequency and scaling factor (η) need to be empirically determined in the image reconstruction process; this prevents HiLo technique to be fully automated, especially when switching among different biological specimens.

In this paper, we present a new two-snapshot SLI reconstruction algorithm for fast image acquisition. The new algorithm, which only requires two mutually $\pi$ phase-shifted raw structured images, is implemented on a custom-designed TFFM to enhance its optical cross-sectioning capability via a digital micromirror device (DMD), which functions as a programmable blazed grating that carries the designed structured patterns. Mathematical models of the new SLI algorithm have been derived; imaging experiments are performed to verify the predicted performance. The new SLI algorithm can be implemented in any sinusoidally modulated illumination system in a fully automatic, adaptive, and robust fashion that enables high-resolution and high-speed 3D imaging. In the following sections, we report the (1) principle of generating structured patterns in a temporal focusing system via a DMD, (2) theoretical development of the two-snapshot SLI algorithm, and (3) SLI TFFM imaging results on pollen samples.

## 2. METHODS

### 2.1 Generation of temporal focusing and structured patterns via a DMD

Figure 1 illustrates the principle of generating structured images in a TFFM based on a DMD. A DMD consists of several millions of binary micromirrors; each mirror has two stable angular positions, i.e., ±12°, and typically operates at a speed of 4.2 – 32.5 kHz. As shown in Fig. 1(a), the femtosecond laser beam is first spatially dispersed by the DMD, which functions as a programmable blazed grating, a lens then collimates these beams, and lastly the objective lens recombines the beams in the focal region. As shown in Fig. 1(b), the dispersion causes the laser pulses to be broadened after the DMD and everywhere before and after the front focal plane of the objective due to spectral-temporal relationship. Temporal focusing is thus achieved at the focal region, where all frequency components spatially overlap, leading to the shortest laser pulse and the capability of optical cross-sectioning [9, 15]. Note that to ensure high diffraction efficiency, the incident angle of the laser beam and the DMD micromirror array should satisfy the blazing condition.

To encode sinusoidal fringe patterns to the image, one may recall the inverse Fourier transform of the sinusoidal fringes is a set of spatial frequencies of opposite polarity, i.e., parallel stripes on the DMD. Accordingly, the sinusoidal fringes can be generated at the focal region by spatially selecting the 0[th] and ±1[st] order diffractions, which form the periodic fringes at the front focal plane of the objective lens, as shown in Fig. 1(c). Observing the results in Figs. 1(b) and 1(c), one may find the plane of temporal focusing and sinusoidal fringes coincide, and accordingly structured images encoded with the desired fringe patterns can be directly generated by controlling the dimension of the "parallel stripes" on the DMD. In addition, since out-of-focus fluorescent signals do not contain the periodic fringe patterns, we can computationally remove the out-of-focus emissions and demodulate the sinusoidal fringe patterns from the in-focus images to enhance the axial resolution of the temporal focusing images. Next, we present the new two-snapshot SLI method and its application for in-focus signal decoding.

### 2.2 Fast two-snapshot SLI for in-focus emission demodulation

We first mathematically describe the imaging field. The image captured by the camera, $D(\vec{r})$, can be expressed as:

$$D(\vec{r}) = S(\vec{r})\{I_0(\vec{r})[1+m\cos(2\pi\vec{p}_\theta \cdot \vec{r}+\varphi)]\}^2 \otimes H(\vec{r})+D_{p2}(\vec{r}) \tag{1}$$

where $\vec{r} \coloneqq (x, y)$ is a 2D spatial position vector; $S(\vec{r})$ represents the fluorophore density distribution within the illuminated sample; $I_0(\vec{r})[1 + m\cos(2\pi\vec{p}_\theta \cdot \vec{r} + \varphi)]$ is a mathematical expression of the intensity distribution of the 0[th] and ±1[st] order beam interference that generates the sinusoidal fringe pattern; $m$ is modulation factor; $\vec{p}_\theta = (|\vec{p}_\theta|\cos\theta, |\vec{p}_\theta|\sin\theta)$ is a frequency vector in the reciprocal space; $\varphi$ is the initial phase of the fringes; $H(\vec{r})$ is the point spread function (PSF) of the imaging system; and $D_{p2}(\vec{r})$ represents the out-of-focus fluorescent signals. Note that time variation for the femtosecond laser pulse is not considered in the analysis. For a two-photon imaging system, the in-focus image is quadratically proportional to the excitation intensity, and thus the first term in $D(\vec{r})$ is squared. The second term in $D(\vec{r})$, i.e., the out-of-focus emissions, does not have frequency modulation [26].

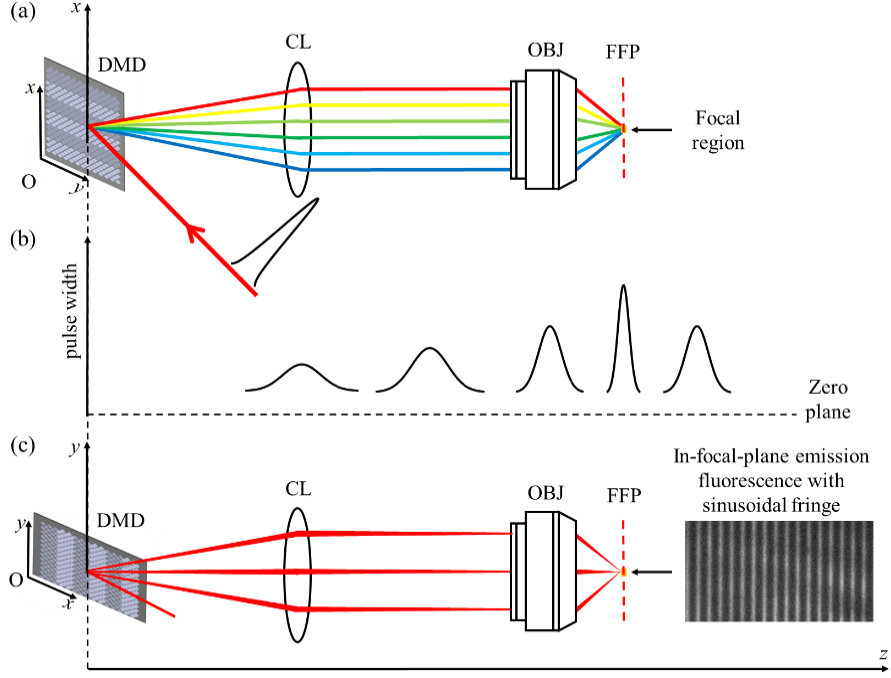

Figure 1. Principles of generating structured images in a TFFM based on a DMD: (*a*) DMD disperses the laser pulses along the optical path until the objective lens (OBJ) recombines the spectral components, achieving temporal focusing at the focal region, where the shortest pulse is formed; CL: collimating lens; FFP: front focal plane; (*b*) illustration of pulse widening effect outside the focal region due to temporal focusing; (*c*) sinusoidal fringe patterns are formed in the focal plane by the interference of the 0$^{th}$ and ±1$^{st}$ order diffractions; the characteristics of the fringes can be controlled by the vertical stripes on the DMD. The right image shows measured in-focus fluorescence emissions encoded with sinusoidal patterns. Note that in (a) and (c), the CL and OBJ form a 4-f system.

To derive the two-snapshot algorithm, first the orientation of the modulated fringe patterns is automatically identified via spatial frequency vector detection. By rotation and subtraction, the modulated in-focus images are obtained. Lastly, a parallel amplitude demodulation method, derived based on Hilbert transform, is applied to complete the decoding processes via two phase-shifted structured images, i.e., $D_1(\vec{r})$ and $D_2(\vec{r})$. The process flow is illustrated in Fig. 2(a). In the following analysis, the phase difference of $D_1(\vec{r})$ and $D_2(\vec{r})$ is set to $\pi$ to simplify the process, i.e., $\varphi_1 - \varphi_2 = (2l + 1)\pi$, where $l$ is an integer. First, we consider the frequency vector, $\vec{p}_\theta$, in the reciprocal space. The position of $\vec{p}_\theta$ can be determined in the reciprocal space by applying local maxima detection. This will allow us to identify the skew angle $\theta$ of the fringe pattern, as shown in Fig. 2(c), by using the formula $\theta = \arctan(p_v/p_u)$, where $(p_v, p_u)$ are the coordinates of $\vec{p}_\theta$, as illustrated in Fig. 2(e). Next, we perform a planar rotation to ensure the fringe pattern is parallel to the y-axis, as shown in Fig. 2(f); the angle of rotation is set to $\pi/2 - \theta$. Image padding processes are applied before the rotation step in order to avoid introducing artefacts in the reconstructed images. The rotated structured images can be mathematically expressed as:

$$D_n^{'}(\vec{r}) = S^{'}(\vec{r})I_0^{'2}(\vec{r})\{1 + \frac{m^2}{2} + 2m\cos(2\pi|\vec{p}_\theta|\cdot x + \varphi_n) + \frac{m^2}{2}\cos[2(2\pi|\vec{p}_\theta|\cdot x + \varphi_n)]\} \otimes H(\vec{r}) + D_{p2}^{'}(\vec{r}) \quad (2)$$

where $n = 1$ or 2; $S'(\vec{r})I_0^{'2}(\vec{r})$ is the rotated in-focus fluorescent signals; and $D_{p2}'(\vec{r})$ is the rotated out-of-focus fluorescent signals. Accordingly, the non-modulated term, $D_{p2}'(\vec{r})$ can be removed by a subtraction step, i.e., $D_d'(\vec{r}) = D_1'(\vec{r}) - D_2'(\vec{r})$. The result is expressed in Eq. (3):

$$D_d^{'}(\vec{r}) \propto S^{'}(\vec{r})[I_0^{'2}(\vec{r})\cos(2\pi|\vec{p}_\theta|\cdot x + \varphi_1)] \otimes H(\vec{r}) \quad (3)$$

where $D_d'(\vec{r})$ is the subtracted image, as illustrated in Fig. 2(g). Note that $D_d'(\vec{r})$ represents the sinusoidally modulated in-focus signal, where its amplitude, $S'(\vec{r})I_0^{'2}(\vec{r})$, is our target of decoding.

Next, a set of complex vectors is constructed based on $D_d'(\vec{r})$ for decoding. First, the 2D image, $D_d'(\vec{r})$, is decomposed to many row vectors, i.e., $V_1(\vec{e}_1), V_2(\vec{e}_2), \ldots, V_k(\vec{e}_k), \ldots, V_s(\vec{e}_s)$, where $s$ represents the total number of rows of $D_d'(\vec{r})$;

$V_k(\vec{e}_k)$ represents the $k^{\text{th}}$ row vector ($k = 1, 2, 3, \ldots, s$); and $\vec{e}_1, \vec{e}_2, \ldots, \vec{e}_k, \ldots, \vec{e}_s$ are the orthogonal bases for $D'_d(\vec{r})$. $V_k(\vec{e}_k)$ is mathematically expressed as:

$$V_k(\vec{e}_k) \propto S'(\vec{e}_k)[I_0^{'2}(\vec{e}_k)\cos(2\pi|\vec{p}_\theta|\cdot x+\varphi_n)]\otimes H(\vec{r}) \tag{4}$$

Accordingly, we can process the 1D complex vectors in a parallel fashion in the computer. To achieve this, we define $VA_k(\vec{e}_k)$ as:

$$VA_k(\vec{e}_k) = V_k(\vec{e}_k) + i\cdot VH_k(\vec{e}_k) \tag{5}$$

where the real part is the $k^{\text{th}}$ row vector, $V_k(\vec{e}_k)$; the imaginary part, $VH_k(\vec{e}_k)$, is generated by the 1D Hilbert transform, i.e., $VH_k(\vec{e}_k) = H\{V_k(\vec{e}_k)\}$ [27]. $VH_k(\vec{e}_k)$ can be mathematically express as:

$$VH_k(\vec{e}_k) \propto S'(\vec{e}_k)I_0^{'2}(\vec{e}_k)\sin(2\pi|\vec{p}_\theta|\cdot x+\varphi_n)]\otimes H(\vec{r}) \tag{6}$$

Next, we parallelly perform amplitude demodulation by recombining the Euclidean norms of the complex vectors. First, substitute the expressions of $V_k(\vec{e}_k)$ and $VH_k(\vec{e}_k)$ into Eq. (5), we have:

$$VA_k(\vec{e}_k) \propto S'(\vec{e}_k)I_0^{'2}(\vec{e}_k)[\cos(2\pi|\vec{p}_\theta|\cdot x+\varphi_n)+i\cdot\sin(2\pi|\vec{p}_\theta|\cdot x+\varphi_n)]\otimes H(\vec{r}) \tag{7}$$

By taking the Euclidean norm of the complex vectors, we arrive at:

$$|VA_k(\vec{e}_k)| \propto S'(\vec{e}_k)I_0^{'2}(\vec{e}_k)\otimes H(\vec{r}) \tag{8}$$

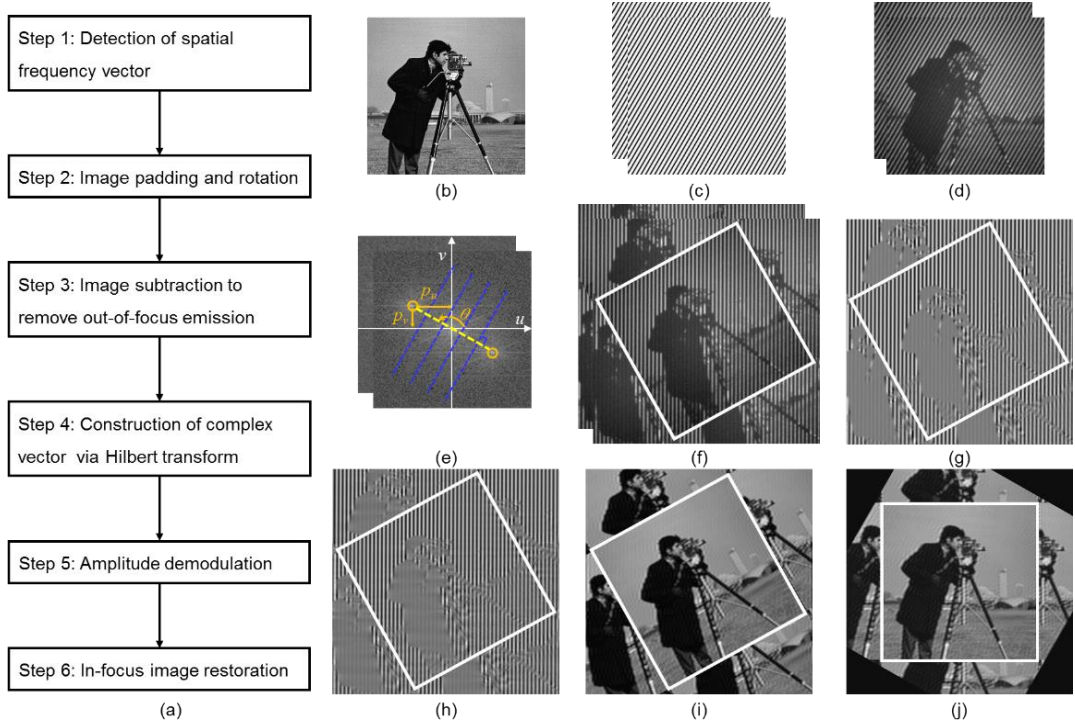

Figure 2. Processing steps of the two-snapshot SLI algorithm: (a) flowchart of the two-snapshot SLI algorithm; (b) test object: image of the cameraman, i.e., $S(\vec{r})$; (c) structured fringe pattern, $1 + m\cos(2\pi\vec{p}_\theta\cdot\vec{r} + \varphi_n)$, $n$ =1 or 2; (d) generation of the raw structured image described in Eq. (1), i.e., the image is simulated by convolving the modulated test object image with the PSF of the optical system; (e) identification of the skew angle, $\theta$, of the spatial frequency vector, $\vec{p}_\theta$, in reciprocal space via local maxima detection; (f) rotation of the structured images by $\pi/2 - \theta$ to align the fringe pattern parallel to the y-axis, described in Eq. (2); image padding is performed before rotation to avoid the generation of artifacts. The region of interest is indicated by the white box; (g) image subtraction to remove the $D'_{p2}(\vec{r})$ term, described in Eq. (3); (h) result of parallel Hilbert transform, where $s$ 1D vectors are reshaped as a 2D image; (i) amplitude demodulation result for the rotated in-focus image, described in Eq. (9); and (j) reconstructed in-focus image, i.e., $S(\vec{r})I_0^2(\vec{r})$.

Importantly, from Eq. (8), $|VA_k(\vec{e}_k)|$ is proportional to $S'_k(\vec{e}_k)I'^2_0(\vec{e}_k)\otimes H(\vec{r})$, where all the sine and cosine modulation terms in the expression are removed. Thus, just by combining $|VH_1(\vec{e}_1)|$, $|VH_2(\vec{e}_2)|$, …, $|VH_k(\vec{e}_k)|$, …., $|VH_s(\vec{e}_s)|$, the rotated in-focus image can be decoded and expressed as:

$$S'(\vec{r})I'^2_0(\vec{r}) \propto [|VH_1(\vec{e}_1)| \; |VH_2(\vec{e}_2)|, ..., |VH_s(\vec{e}_s)|]^T \tag{1}$$

The decoded in-focus image is illustrated in Fig. 2(i). To fully restore the in-focus fluorescent signals, $S(\vec{r})I^2_0(\vec{r})$, we lastly rotate $S'(\vec{r})I'^2_0(\vec{r})$ by an angle of $\theta - \pi/2$ to align the resultant image to its original orientation, as shown in Fig. 2(j). Note that since the fringes are parallel to the *y*-axis, it is unnecessary for our method to implement the bidimensional empirical mode decomposition (BEMD) and spiral phase Hilbert transform [28] in the amplitude demodulation step, making the algorithm simple and efficient. It is worthwhile to note that if the optical system is perfectly aligned such that the modulation fringes are parallel to the *y*-axis, step 1 and step 2, i.e., detection of spatial vector as well as image padding and rotation, may be entirely omitted, leading to shorter calculation time.

## 3. EXPERIMENTAL RESULTS

### 3.1 Experimental setup

Figure 3 presents the optical configuration of the DMD-based TFFM. The laser source is a Ti:sapphire regenerative laser amplifier (Spitfire Pro, Spectra-Physics) with an average power of 4W; pulse width of 100 fs; repetition rate of 1 kHz; center wavelength of 800 nm; and output beam diameter of ~12 mm. A mechanical shutter is included in the system to control the excitation time. A half-wave plate and a polarizing beam splitter together control the laser power. Two high reflectance mirrors, M1 and M2, are used to guide the laser beam to the DMD (DLP 4500, 912 × 1140 pixels; pixel size: 7.6 × 7.6 μm, Texas Instrument), where the DMD replaces the blazed grating in a generic temporal focusing system and simultaneously functions as a diffraction grating and a programmable binary mask. To achieve high efficiency, the incident angle of the laser beam is set to 24.8° in reference to the DMD surface so that the collected 5th order diffraction satisfies the blazing condition. After the collimating lens (CL), an objective lens (CFI Apo LWD 40X; NA 1.15, Nikon) recombines the dispersed spectral components on the x-z plane, shown in Fig. 1(a); and temporal focusing is achieved at the focal plane of the objective lens.

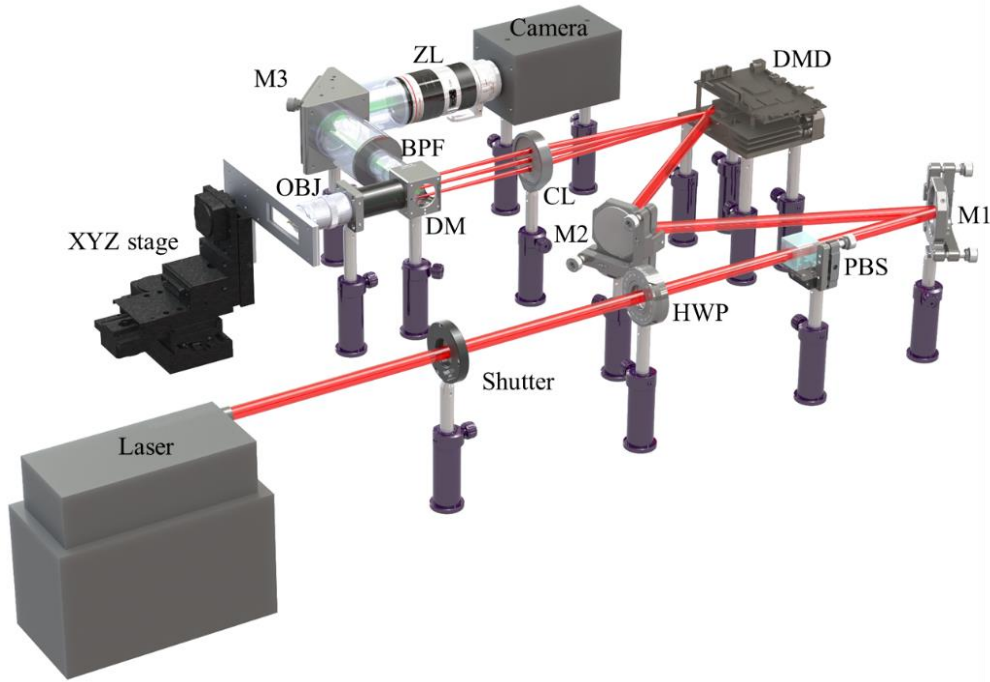

Figure 3. Schematics of the DMD-based SLI TFFM system. Laser: regenerative laser amplifier; HWP: half-wave plate; PBS: polarizing beam splitter; M1 - M3: mirrors; CL: collimating lens; DM: dichroic mirror; BPF: band pass filter; OBJ: objective lens; ZL: zoom lens.

Note that the objective lens simultaneously recombines the 0$^{th}$ and ±1$^{st}$ order diffractions on the y-z plane, shown in Fig. 1(c) and Fig. 3, to form the sinusoidal fringes at the focal plane. The field of view is ~300 × 250 µm$^2$, where each DMD pixel maps to an area of 270 × 270 nm$^2$. The sample is mounted on a precision xyz stage (MP-285, Sutter Instrument) for positioning. The emissions are collected by the objective lens, separated from the excitation signals via a dichroic mirror (ET780lp, Chroma) and a band-pass filter, and lastly imaged to a high sensitivity electron-multiplying charge-coupled device (EMCCD) camera (ProEM, Princeton Instrument, USA). A camera zoom lens (EF 70-200 mm f/2.8L IS II USM, Canon) is installed before the EMCCD camera for the ease of adjusting the focal plane. A data acquisition (DAQ) card (USB-6363, National Instruments) collects and processes the signals to form images.

### 3.2 Generation of sinusoidal fringes via the DMD

Before implementing the two-snapshot algorithm, we first verify the DMD can precisely generate and control the designed sinusoidal fringe patterns. A thin layer of Rhodamine 6G (Rh6G) mixed in polymethyl-methacrylate (PMMA) is prepared [29] to illustrate the capability of the DMD. In the experiment, the on-off states of the DMD micromirrors are controlled electronically to generate parallel stripes of 50% duty cycle, as discussed in Section 2.1. The period of the stripe pattern is set as 2.70 µm, i.e., 10 DMD pixels. Note that $\pi$-phase shifted patterns can be generated by applying binary patterns of reversed states. Figures 4(a) and 4(b) present the binary patterns on the DMD that have opposite states; the imaging results of the Rh6G sample are presented in Figs. 4(c) and 4(d), respectively. The fluorescent intensity profiles along the red and blue lines in Figs. 4(c) and 4(d) are plotted in Fig. 4(e), where the circles represent the measured fluorescent intensities, and solid lines represent the least-squares fitted results using $[1+\cos(|\vec{p}_\theta|\cdot x+\varphi_n)]^2$, where $n$ = 1 or 2. From the results, one can conclude that the measured intensities have sinusoidal profiles; and the two curves in Fig. 4(e) have a precise phase shift of 180°. Figure 4(f) presents the Fourier transform results of the measured intensity data in Fig. 4(e), where one can clearly observe the color-coded ±1$^{st}$ peaks have opposite signs, confirming again the 180° phase shift. Importantly, from Fig. 4(f), we also observe the 2$^{nd}$ order peaks, as indicated by the black arrows, which represent the second harmonic frequency component, i.e., the $\frac{m^2}{2}\cos[2(2\pi|\vec{p}_\theta|\cdot x + \varphi_n)]$ term in Eq. (2).

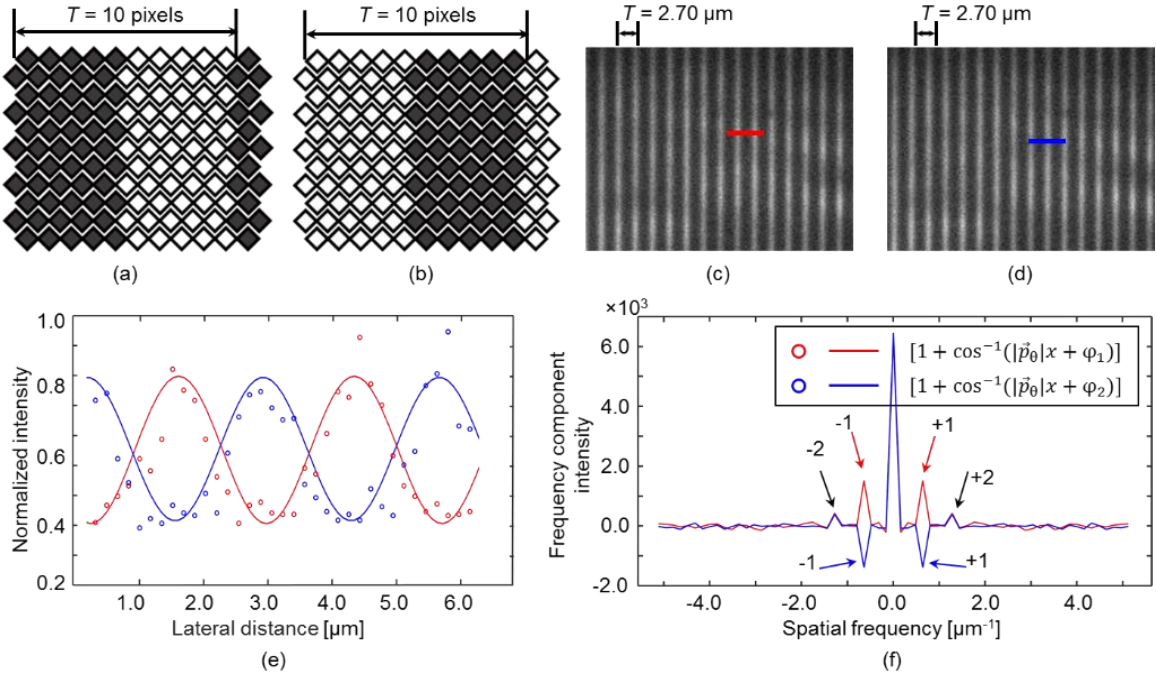

Figure 4. Generation of $\pi$ phase-shifted sinusoidal fringe patterns via a DMD. (a) and (b): binary fringe patterns programmed to the DMD of opposite states; the period of the stripe pattern is ten pixels (2.70 µm); (c) and (d): imaging results of the Rh6G sample at the focal plane when applying DMD patterns of (a) and (b) respectively; (e) measured fluorescent intensity profiles across the red and blue lines in (c) and (d) respectively. The circles and solid lines represent the raw data and the least-squares fitting results, which have a 180° phase shift; (f) Fourier transform of the fluorescent intensity profiles in (e), where the 1$^{st}$ and 2$^{nd}$ order peaks can be clearly observed.

### 3.3 Axial resolution enhancement

In this section, we experimentally characterize the axial resolution of the SLI TFFM based on the two-snapshot algorithm via the thin Rh6G sample. The axial resolution is determined by axially scanning the Rh6G sample with 1 μm steps over a range of 20 μm, where the fluorescent intensities are recorded at every step. The results are presented in Fig. 5, where the red and blue circles represent the measured fluorescent intensities of the TFFM "without" and "with" the two-snapshot algorithm respectively. The solid lines are the least-squares fitted curves of a Gaussian. Analyzing the results, one can conclude that the axial resolution, i.e., full width at half maximum (FWHM), of the TFFM "without" and "with" the two-snapshot algorithm is 5.84 μm and 4.30 μm respectively. Accordingly, the axial resolution enhancement factor, defined in [30], is calculated to be 1.25.

Note that the theoretical prediction of the axial resolution of a TFFM is derived and reported in [31,32]. By substituting our system characteristics, i.e., *NA* of objective = 1.15, refractive index of the immersion fluid $n$ = 1.33; and magnification of the system $M$ = 40, into the equation, the optimal axial resolution of our system is calculated to be 3.79 μm. Next, based on Stokseth's approximation [33], we can calculate the axial resolution for structured light illumination systems. By substituting the system characteristics again, the axial resolution is found to be 2.73 μm. Accordingly, the predicted axial resolution enhancement factor is 1.27, which matches well with the experimental result. It is worthwhile to note that the theoretical axial resolution of the two-snapshot algorithm is identical to the conventional RMS algorithm; in other words, the two-snapshot algorithm improves the data acquisition speed by a factor of 2.5 without sacrificing the image resolution.

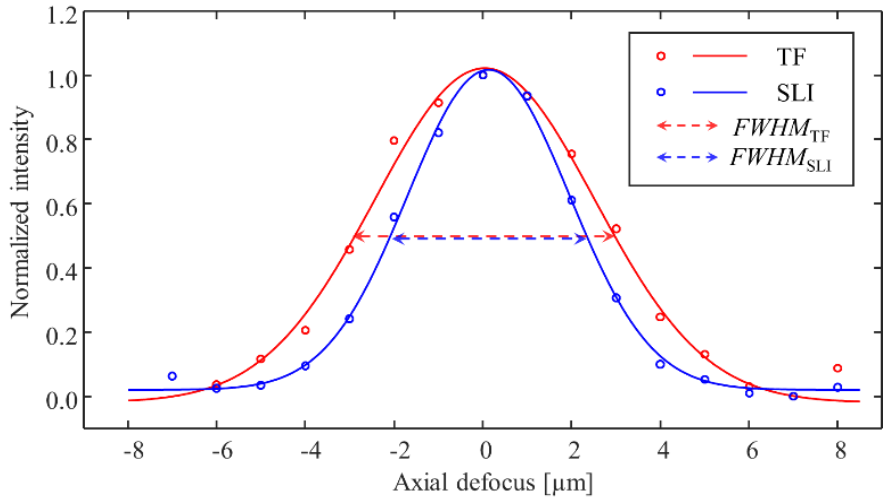

Figure 5. Axial resolution characterization via the thin Rh6G sample; the red and blue circles represent the measured fluorescent intensities of the TFFM "without" and "with" the two-snapshot algorithm.

### 3.4 Imaging experiments on pollen grain samples

Imaging experiments on pollen grain samples (Item# 304264, Carolina Biological Supply) have been performed to characterize the performance of the TFFM with the two-snapshot algorithm. Figures 6(a) - 6(c) present the imaging results of a pollen grain (~30 × 30 μm$^2$, cropped from the original image) at three different depths, i.e., 0 μm, 6 μm, and 12 μm without SLI algorithms. From the images, one can observe the presence of out-of-focus fluorescent signals, which blur the image. Figures 6(d) - 6(f) present the reconstructed images via two snapshots in the same imaging field. From the results, we find the out-of-focus emissions have been effectively suppressed by the two-snapshot algorithm; and low contrast features become evident and clear, i.e., improved signal to background ratio (SBR). The effect may be clearly observed when one compares the contours of the pollen in Figs. 6(c) and 6(f). Figure 6(g) presents normalized intensity profiles along the green and blue dashed line in Figs. 6(c) and 6(f) respectively. The results confirm the strong reduction in mean background intensity, i.e., from 0.61 to 0.32 (calculated along the dashed lines), and an increase of SBR value from 1.4 to 2.3. Overall, the imaging results demonstrate the capability of the two-snapshot method on (1) out-of-focus fluorescent emission rejection and (2) axial resolution enhancement.

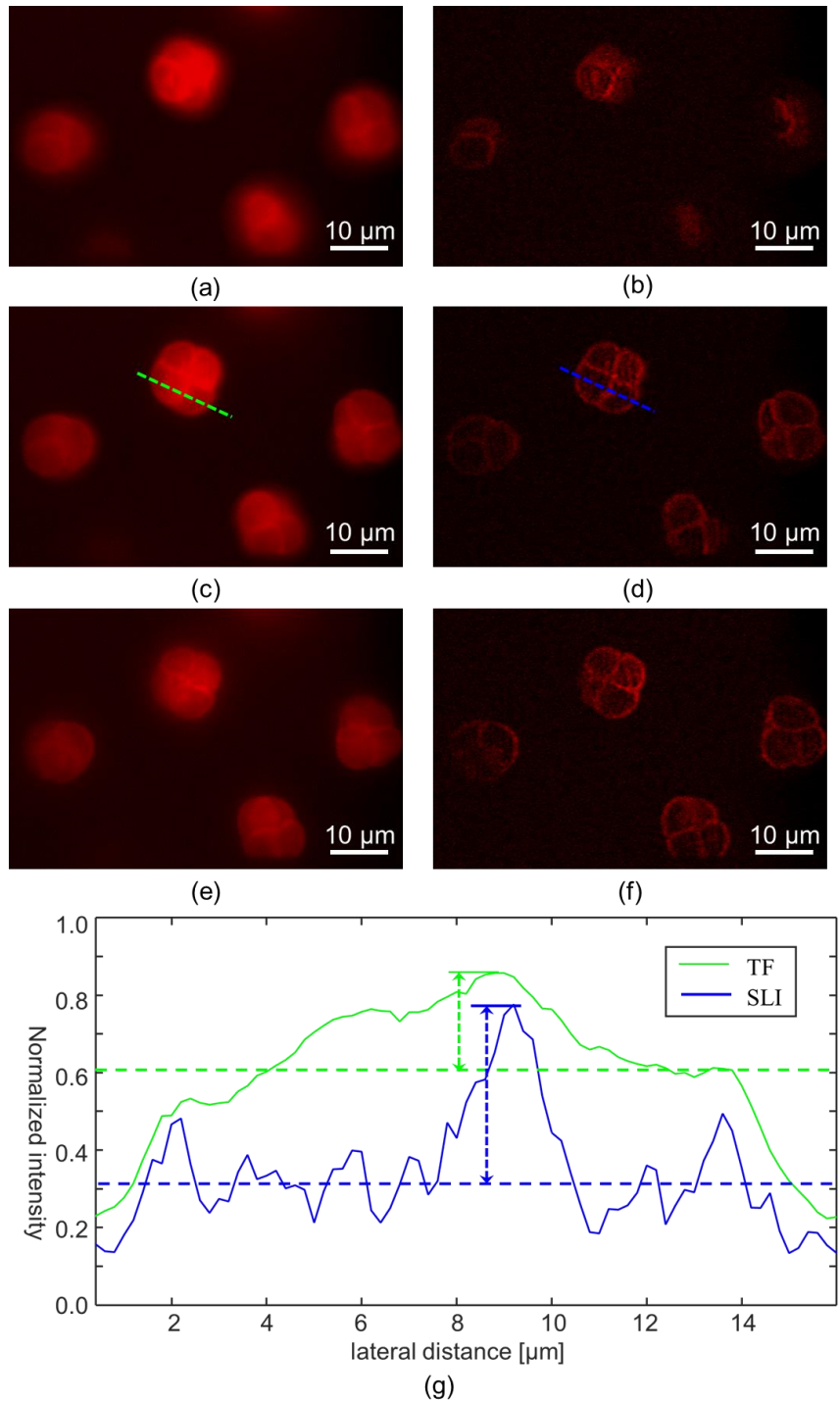

Figure 6. TFFM imaging results of a pollen grain at three different depths. (a) - (c) cropped optical cross-sections of the pollen at 0 µm, 6 µm, and 12 µm respectively; (d) - (f) two snap shots reconstructed images of the same pollen grain at 0 µm, 6 µm, and 12 µm respectively, where one may clearly observe the enhanced axial resolution and effect of out-of-focus fluorescent emission rejection in (c) and (f); (g) normalized intensity profiles along the red and blue dashed line in (c) and (f). The scale bar is 10 µm.

In summary, comparing with the conventional RMS algorithm, the two-snapshot algorithm has the following advantages: (1) the image acquisition speed is increased, i.e., a factor of 2.5 in two-photon microscopy, as only two raw images are required versus five for the RMS method; and (2) at the same imaging speed, the two-snapshot method requires less raw images, resulting in images with better signal-to-noise ratios. One limitation of the two-snapshot algorithm is that the Hilbert transform is a relatively time-consuming operation and thus the reconstructed images may not be displayed in real-time (note the raw images can be displayed at the acquisition speed), which the RMS algorithm may allow when paired with a high-performance computer. This drawback may be addressed by more powerful computers with parallel computing techniques [34]. Comparing with other algorithms that require two raw images, i.e., the HiLo technique [23–25] and BEMD [28], our new two-snapshot algorithm has the advantage of being completely adaptive without the need of manual inspection and selection of critical operation parameters, such as scaling factor, cut-off frequency, or bidimensional intrinsic mode functions etc. Note that the two-snapshot algorithm has superior computation speeds versus both the HiLo method and BEMD method.

## 4. CONCLUSION

We have presented a new two-snapshot SLI algorithm for fast image acquisition that only requires two mutually $\pi$ phase-shifted raw images. The new algorithm is implemented on a custom-built DMD-based TFFM to demonstrate its unique characteristics, including the enhanced axial resolution and background rejection capability. Mathematical models of the two-snapshot method have been derived. The axial resolution enhancement has been experimentally characterized; the results show an enhancement factor of 1.25, which matches well with the theoretical prediction, i.e., 1.27. Imaging experiments have been devised and performed on pollen as well as mouse kidney slices; the experimental results show clean optical cross-sectional images, confirming the great background rejection capability and improved axial resolution. Importantly, these results have been achieved with only two raw images in a completely adaptive and automated fashion, promising 2.5 times higher image acquisition speed. The new two-snapshot algorithm can be readily adapted to any sinusoidally modulated imaging systems, e.g., LED- or continuous wave laser-based SLI microscopes [35, 36], multifocal structured illumination microscope [37], light sheet microscopy [38], or electron microscope, and generate significant impact to the field of biomedical imaging.

## 5. ACKNOWLEDGEMENT

HKSAR Research Grants Council (RGC), General Research Fund (GRF) (CUHK 14202815); Innovation and Technology Commission (ITC), Innovation and Technology Fund (ITF), ITS/007/15P.